\documentstyle[preprint,aps]{revtex}

\begin{document}

%%%%%%%%%%%%%%%%%%%%%%%%%%%%%%%%%%%%%%%%%%%%%%%%%%%%%%%%%%%%%%%%
%%
%%    MACROS  (TO BE REMOVED LATER)
%%
%%%%%%%%%%%%%%%%%%%%%%%%%%%%%%%%%%%%%%%%%%%%%%%%%%%%%%%%%%%%%%%%

\def \CKM{\vert V_{ud} \vert^2}
\def \GeV{{\rm \enspace GeV}}
\def \thetas{\theta^*}
\def \beq{\begin{equation}}
\def \eeq{\end{equation}}
\def \beqa{\begin{eqnarray}}
\def \eeqa{\end{eqnarray}}
\def \amp{{\cal M}}
\def \Wprop{{\cal W}}
\def \up{\uparrow}
\def \down{\downarrow}
\def \half{\hbox{$1\over2$}}
\def \ts{\thinspace}
                
\draft
\preprint{
  \parbox{2in}{Fermilab--Pub--96/417-T \\
  [-0.12in] UM--TH--96--17 \\
  [-0.12in] hep-ph/9611367
}  }

\title{Improved Spin Basis for Angular Correlation Studies \\
in Single Top Quark Production at the Tevatron}
\author{Gregory Mahlon  \cite{GDMemail}}
\address{Department of Physics, University of Michigan \\
500 E. University Ave., Ann Arbor, MI  48109 }
\author{Stephen Parke \cite{SPemail}}
\address{Fermi National Accelerator Laboratory \\
P.O. Box 500, Batavia, IL  60510 }
\date{November 19, 1996}
\maketitle
\begin{abstract}
We show in single top quark production that the spin of the top quark
is correlated with the direction of the $d$-type quark in the event.
For single top production in the  $W^*$ channel, the $d$-type quark 
comes dominantly from the antiproton at the Tevatron, whereas for 
the $W$-gluon fusion channel the spectator jet is the $d$-type quark 
the majority of the time at this machine.  Our results are that
98\% of the top quarks from the $W^{*}$ process have their spins in 
the antiproton direction, and 96\% of the top quarks in the $W$-gluon 
fusion process have their spins in the spectator jet direction.
We also compare with the more traditional, but less effective, 
helicity basis.  The direction of the top quark spin is reflected in 
angular correlations in its decay products.
\end{abstract}
\pacs{}

%%%%%%%%%%%%%%%%%%%%%%%%%%%%%%%%%%%%%%%%%%%%%%%%%%%%%%%%%%%%%%%%
%%
%%      INTRODUCTION
%%
%%%%%%%%%%%%%%%%%%%%%%%%%%%%%%%%%%%%%%%%%%%%%%%%%%%%%%%%%%%%%%%%

\section{Introduction}

The single top quark production processes are of great importance
at hadron colliders
since they allow a direct measurement of the coupling
of the $W$-boson to the top quark
{\it i.e.} the CKM matrix element $|V_{tb}|$.
These processes can also be used to search for anomalous
couplings of the top quark. 
With a mass in the neighborhood of 175 GeV\cite{TopMass},
the top quark is by far the heaviest of the known quarks.
As a consequence, the electroweak decay of the top quark proceeds
so rapidly that 
toponium bound states and $T$ mesons
do not have time to form~\cite{Bigi}
and the decay products of the top quark are correlated
with its spin~\cite{Spin}.
Therefore, if a top quark is produced with a significant spin
correlation, this correlation will be translated into
large angular correlations in such events.
Studies of these angular correlations in single top production
can then be used as sensitive searchs
for anomalous couplings of the top quark,
{\it i.e.}\ physics beyond the Standard
Model~\cite{Kane1,CarlsonIowa,HeinsonIowa,Atwood}.

Traditionally in high energy physics processes,
the discussions of spin-related observables take
place in terms of the helicities of the fermions involved. 
However this description is most useful when the fermions
are produced in the ultra-relativistic limit 
because in this limit the chirality eigenstates are identical
to the helicity eigenstates.
For fermions which are not ultra-relativistic,
such as the top quark produced at the Tevatron,
one must deal with the fact that the chirality and
helicity of a massive fermion may not be specified simultaneously.
Therefore, there is no {\it a priori}
reason to believe that the helicity basis will give the 
best description of the spin of top quarks at the Tevatron.
In fact, it has recently been shown that the helicity basis does
not lead to the largest values for various spin-related correlations
in $t\bar{t}$ production at either the Tevatron\cite{TopPairsHadronic}
or an $e^{+}e^{-}$ collider\cite{TopPairsLeptonic}.  
Thus, it is natural to ask: is there a better spin basis
than helicity for the description of the spin correlations in {\it single} 
top production?  The answer to this question is yes:
we will construct such a spin basis in this paper.

\begin{figure}[h]

\vspace*{15cm}
\includegraphics{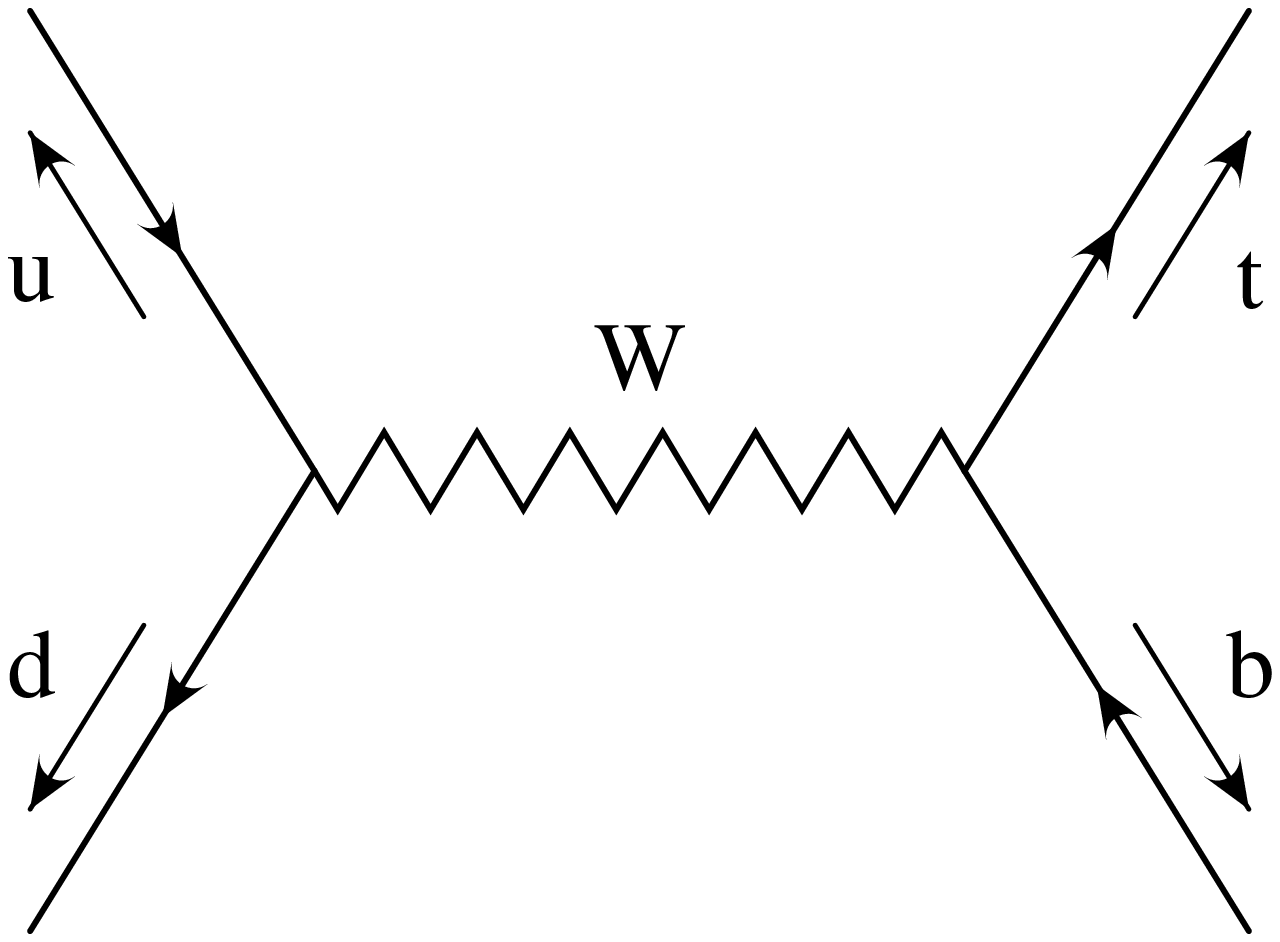}
\vspace{1.0cm}

\caption[]{Feynman diagram for single top production in the $W^*$
process.  The labels indicate the momentum flow utilized in the text.}
\label{WStarDiagram}
\end{figure}

\begin{figure}[h]

\vspace*{15cm}
\includegraphics{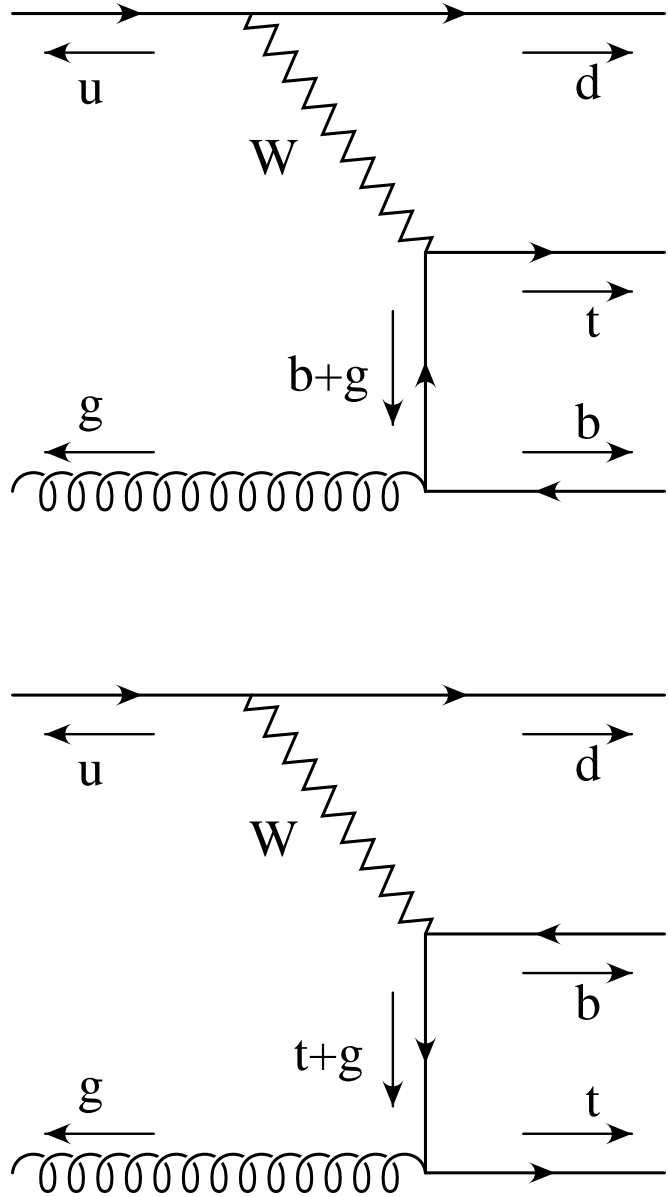}
\vspace{1.0cm}

\caption[]{Gauge-invariant set of Feynman diagrams for 
single top production via $Wg$ fusion.  The labels indicate
the momentum flow utilized in the text.}
\label{WgFusionDiagram}
\end{figure}

We will concentrate on two important production mechanisms for
single top quarks at the Tevatron. 
Both of these mechanisms produce the single top quark in a left handed
chirality state through a virtual $W$ boson. 
Therefore, significant spin correlations are 
expected even at the Tevatron, where the top quark is produced 
well below the ultra-relativistic limit.
The first of these 
(Fig.~\ref{WStarDiagram})
is the purely electroweak $W^{*}$ 
channel~\cite{Cortese,WstarLO,WstarLoop}
\beq
u \bar{d} \rightarrow t\bar{b},
\label{WStarProcess}
\eeq
while the second 
(Fig.~\ref{WgFusionDiagram})
consists of the so-called $W$-gluon fusion ($Wg$ fusion)
processes~\cite{Dawson,WillenbrockDicus,DawsonWillenbrock,Yuan1,EllisParke,UNK,CarlsonPLB}
\beqa
u g && \rightarrow t \bar{b} d;  \cr 
\bar{d} g && \rightarrow t \bar{b} \bar{u}.
\label{WgFusionProcess}
\eeqa 
The feasiblity of isolating single top quark production
in a collider environment has been already been demonstrated
for both the $W^{*}$ channel~\cite{WstarLO,TeVMM} and
the $W$-gluon fusion process~\cite{CarlsonPLB}.

Early on it was recognized by 
Willenbrock and Dicus~\cite{WillenbrockDicus}
that the $Wg$ fusion process is dominated by the configuration
where the $\bar{b}$ quark is nearly collinear with the incoming
gluon, leading to a logarithmic factor $\ln(m_t^2/m_b^2)$
in the total cross section.  In the event that this factor is
too large\footnote{The authors of Ref.~\cite{WillenbrockDicus}
suggest $(g_s^2/4\pi^2)\thinspace\ln(m_t^2/m_b^2) \approx 0.23$
as a suitable measure,
where $g_s$ is the strong coupling constant,
the top quark mass $m_t = 175 \GeV$,
and the bottom quark mass $m_b = 5 \GeV$.}
the perturbative calculation of the 
$2\rightarrow3$ process becomes unreliable, and one should instead
compute $ub\rightarrow td$, with the large logarithm being
absorbed into the $b$ parton distribution function.
This latter approach has been employed by Bordes 
{\it et. al.}~\cite{Bordes1,Bordes2,Bordes3} in their effort 
to accurately compute
the total cross section, including higher-order corrections.
Among their conclusions is the statement that to lowest order,
for top quark masses up to a few hundred GeV, 
the two pictures give comparable
event descriptions and lead to 
similar cross sections~\cite{Bordes3}.  
This statement is also true for the top quark spin correlation.
Therefore, we will frame
our discussion of $Wg$ fusion
in terms of the tree level description involving only the diagrams
in Fig.~\ref{WgFusionDiagram}.

In Sections II and III we discuss in detail the top quark spin
correlations in single top quark production via 
the $W^{*}$ process 
and $Wg$ fusion process, respectively. Finally we end with
a discussion and conclusions. In the Appendix we give an example
in detail of how the top spin correlations 
lead to angular correlations in events.
The example given is single top production in the $W^*$ channel.
Throughout this paper we will only consider processes which produce
a top quark in the final state. 
The treatment of the charge-conjugated processes,
where a top antiquark is produced,
is similar.

%%%%%%%%%%%%%%%%%%%%%%%%%%%%%%%%%%%%%%%%%%%%%%%%%%%%%%%%%%%%%%%%%
%%
%%      W-star PRODUCTION MECHANISM
%%
%%%%%%%%%%%%%%%%%%%%%%%%%%%%%%%%%%%%%%%%%%%%%%%%%%%%%%%%%%%%%%%%

\section{Single Top Production Through A $W^{*}$} \label{WStarSection}

We begin with the simpler of the two
production mechanisms for single top quarks at the
Tevatron, the electroweak process $u\bar{d}\rightarrow t\bar{b}$,
which proceeds via a virtual $W$ boson (see Fig.~\ref{WStarDiagram}).
We represent the momentum of the each particle by its symbol,
and write the amplitude in crossing symmetric form with all
momenta outgoing.
Our results are easily derived
using the spinor helicity method for massive fermions described
in~\cite{TopPairsHadronic} to treat the top spin.  In particular,
we decompose the top quark momentum into a sum of two
massless auxiliary
momenta, 
\beq
t_{1} \equiv \half(t + m_ts); \qquad
t_{2} \equiv \half(t - m_ts),
\eeq
where $s$ is
the usual spin vector of the top quark.  
In the rest frame of the top quark, the spin of the top quark
is in the same direction as the spatial part of $t_1$.
Then, the matrix element squared for the production of
a spin up top quark summed over color and all of the other 
spins\footnote{Although we have summed over the spins and colors of the
initial particles, we have not performed the spin or color
average in any of the
matrix elements appearing in this paper.}
is
\beq
\vert\amp(0\rightarrow \bar{u} d t_{\up}\bar{b})\vert^2
=
g_W^4 \CKM N_c^2 \thinspace
{
{ (2d\cdot t_2)(2u\cdot b) }
\over
{ (2u\cdot d-m_W^2)^2 + (m_W\Gamma_W)^2 }
}
\label{WStarGenUP}
\eeq
while for a spin down top quark we have
\beq
\vert\amp(0\rightarrow \bar{u} d t_{\down}\bar{b})\vert^2
=
g_W^4 \CKM N_c^2 \thinspace
{
{ (2d\cdot t_1)(2u\cdot b) }
\over
{ (2u\cdot d-m_W^2)^2 + (m_W\Gamma_W)^2 }
},
\label{WStarGenDN}
\eeq
where $g_W$ is the weak coupling constant, $m_W$ and $\Gamma_W$ are
the mass and width of the $W$ boson, $N_c$ is the number
of colors, and $V_{ud}$ is the Cabibbo--Kobayashi--Maskawa
matrix element.
Throughout this paper we assume the Standard
Model with three generations and suppress the CKM factor
$\vert V_{tb}\vert^2 \approx 1$.
The sum of~(\ref{WStarGenUP}) and~(\ref{WStarGenDN}) is
obviously independent of the choice of the spin axis 
of the top quark, as is required.

It is clear that the top quarks
produced via the $W^*$ process
are 100\% polarized along the direction
of the $d$-type quark, since~(\ref{WStarGenDN}) vanishes if we
choose $t_1\propto d$.  Consequently, the ideal basis for 
studying the $t$ spin is the one which uses 
the direction of the $d$-type quark as the spin axis.
(See the Appendix for a discussion of this process 
keeping track of all of the correlations between production
and decay.)
Of course, in an actual experiment, we know only that one
of the two initial state partons is a $\bar{d}$.
However,  the largest contribution to the total cross section 
comes from the case where the $\bar{d}$ is donated 
by the antiproton.  In fact, for the
Tevatron at 2~TeV, we estimate that 98\% of the cross section 
may be attributed to this configuration 
(see Table~\ref{WStarPartons}).  
This suggests that an excellent 
choice would be to decompose the top spin along the direction of 
the antiproton beam, independent
of the actual identity of the parton supplied by that beam.
We will refer to this choice as the ``antiproton'' basis.

To aid in the comparison of the antiproton basis to the
more traditional helicity decomposition, we present
the matrix elements as a function of 
the angle $\thetas$ between the direction
of the top quark in the zero momentum frame (ZMF) 
of the initial parton pair and the
${+}z$ axis, and the speed $\beta$ of the top quark
in the ZMF.  We orient our coordinate system such that 
the protons travel in the positive $z$ direction; 
the antiprotons travel in the negative $z$ direction.
Because Eqs.~(\ref{WStarGenUP}) and~(\ref{WStarGenDN})
are not symmetric under the interchange of $u$ and $d$,
the expressions we obtain in terms of these variables
will depend upon which beam the $\bar{d}$ quark comes from.
In the following equations, the parton taken from the proton
will always be written first, followed by the parton taken
from the antiproton.

We now turn to the actual matrix elements for the antiproton
basis, where spin up means that in the rest frame of the top
quark, its spin points in the same direction as the incoming
antiproton beam is traveling in that frame.
For the 98\% of the time that the $\bar{d}$ comes from
the antiproton, we have 
\beq
\vert\amp(u\bar{d}\rightarrow t_{\up}\bar{b})\vert^2
=
{ {g_W^4 \CKM N_c^2}\over{\Wprop} }
\thinspace\beta(1+\cos\thetas)(1+\beta\cos\thetas),
\label{WStarPbarUP}
\eeq
where
\beq
\Wprop \equiv
\biggl[(1 - \xi) + \beta(1+\xi)\biggr]^2
+\biggl[ \xi(1-\beta){{\Gamma_W}\over{m_W}}\biggr]^2
\eeq
and $\xi\equiv m_W^2/m_t^2$.
The spin down amplitude vanishes in this case.  On the
other hand, when the $\bar{d}$ is supplied by the proton
instead, %$t_1 = u$ and 
we obtain
\beq
\vert\amp(\bar{d}u\rightarrow t_{\up}\bar{b})\vert^2
=
{ {g_W^4 \CKM N_c^2}\over{\Wprop} }
\thinspace
{
{ \beta^3(1-\cos^2\thetas)(1-\cos\thetas) }
\over
{ 1+\beta\cos\thetas }
}
\label{WStarPbarPU}
\eeq
for spin up, and
\beq
\vert\amp(\bar{d}u\rightarrow t_{\down}\bar{b})\vert^2
=
{ {g_W^4 \CKM N_c^2}\over{\Wprop} }
\thinspace
{
{ \beta(1-\beta^2)(1-\cos\thetas) }
\over
{ 1+\beta\cos\thetas }
}
\label{WStarPbarND}
\eeq
for spin down.  Thus, the presence of a $\bar{d}$ sea 
in the proton introduces
a small quantity of spin down top quarks into the sample.
Indeed, this contribution is dominated by the spin down
component.  However, the smallness of this contribution still
results in a sample in which the top spin is aligned with 
the antiproton direction 
in the top quark rest frame 98\% of the time.

We now compare these results to the 
helicity basis, using the ZMF as the frame in
which we measure the helicity.
We begin with the case where the $\bar{d}$ quark comes
from the antiproton,
for which the matrix element squared is
\beq
\vert\amp(u\bar{d}\rightarrow t_{\up}\bar{b})\vert^2
=
{1\over2}\thinspace{ {g_W^4 \CKM N_c^2}\over{\Wprop} }
\thinspace\beta(1-\beta)(1-\cos^2\thetas)
\label{WStarHelUP}
\eeq
for the production of spin up (right-handed helicity) top quarks and
\beq
\vert\amp(u\bar{d}\rightarrow t_{\down}\bar{b})\vert^2
=
{1\over2}\thinspace{ {g_W^4 \CKM N_c^2}\over{\Wprop} }
\thinspace\beta(1+\beta)(1+\cos\thetas)^2
\label{WStarHelDN}
\eeq
for the production of spin down top quarks.
The expressions
for the $\bar{d}u$ intial state
may be obtained by making the replacement
$\cos\thetas \rightarrow - \cos\thetas$.  
The spin up amplitude is proportional 
to $1{-}\beta$, causing
it to vanish in the ultra-relativistic limit. 
At more moderate values of $\beta$, such as
are dominant at the Tevatron, both spins are produced, with
spin down (left-handed helicity)
top quarks predominating.
We find that in the over-all mixture 
at the Tevatron, 83\% of the top quarks have left-handed helicity.

\begin{figure}[h]

\vspace*{16cm}
\includegraphics{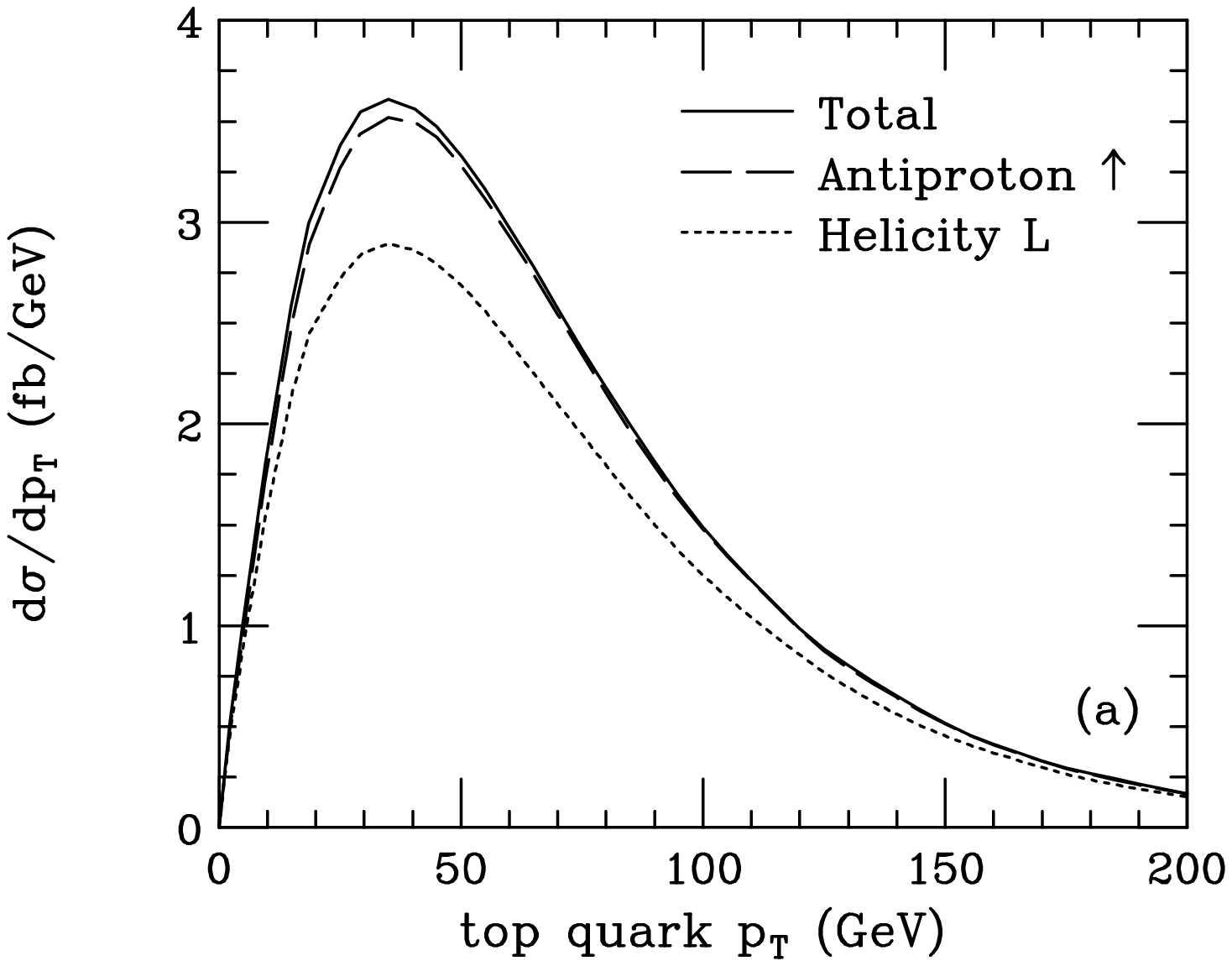}
\includegraphics{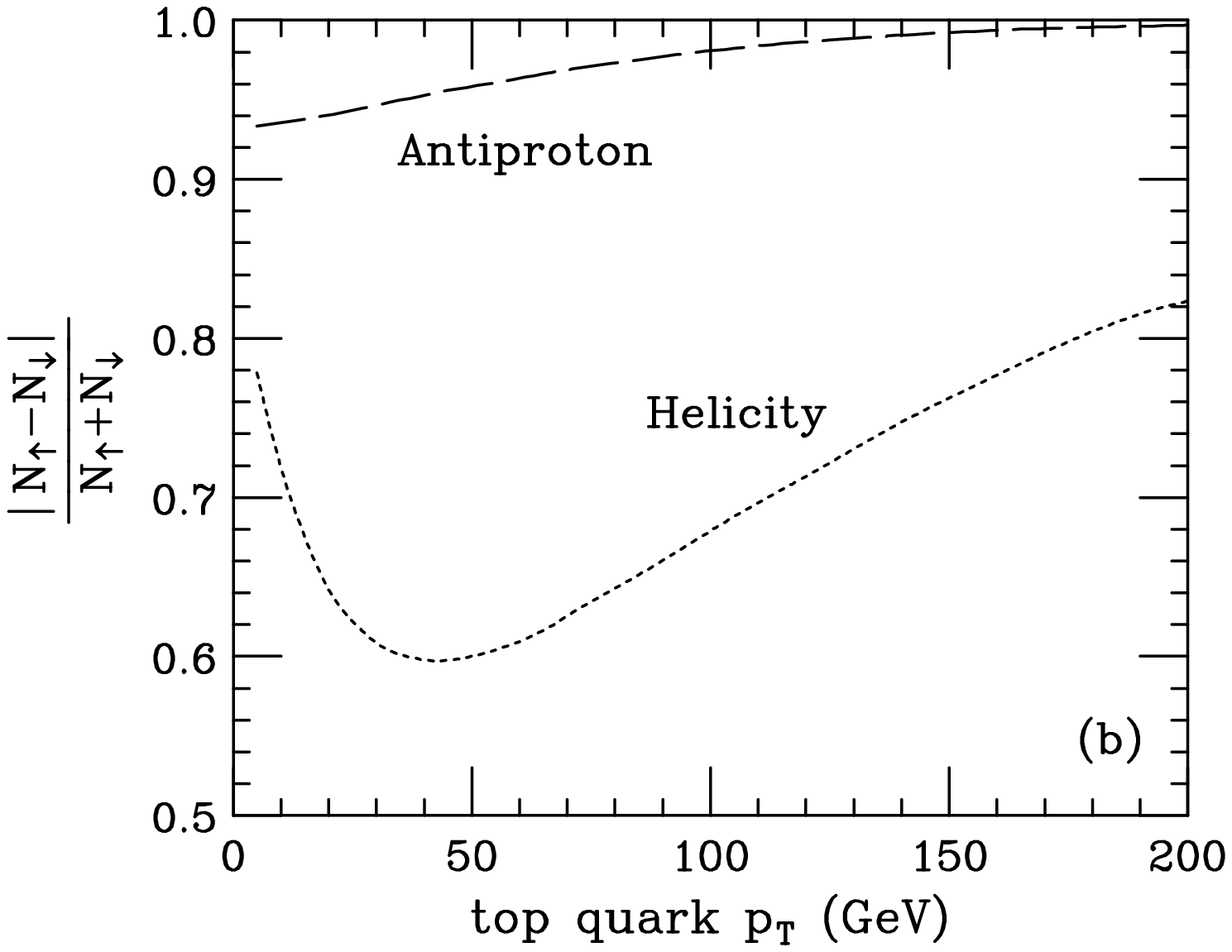}
\vspace{1.0cm}

\caption[]{(a)  The differential cross sections
(total, antiproton basis spin up, helicity basis left)
as a function
of the top quark transverse momentum for single top 
production in the $W^*$ channel at the Tevatron at 2.0 TeV.
(b)  The absolute value of
the spin asymmetry~(\ref{Asym}) plotted as a function of 
the top quark
transverse momentum for the helicity and antiproton bases.
}
\label{WStarPlot}
\end{figure}

Table~\ref{WStarPurities} summarizes the purities for the
helicity, antiproton, and for completeness, proton bases.
The proton basis is defined by choosing $t_1 \propto p$,
{\it i.e.}\ the parton donated by the proton 
beam.\footnote{The analytic forms of the matrix elements
squared in the proton basis are easily obtained:
Eqs.~(\ref{WStarPbarPU}) and~(\ref{WStarPbarND}) apply
to the $u\bar{d}$ initial state once the replacement
$\cos\thetas \rightarrow {-}\cos\thetas$ is made.
Likewise, Eq.~(\ref{WStarPbarUP}) represents the lone non-vanishing
contribution from the $\bar{d}u$ initial state after
making the same replacement.}
Also included are the values of the spin asymmetries
\beq
{{{N_{\up}-N_{\down}}
\over{N_{\up}+N_{\down}}}}
\label{Asym}
\eeq
for each basis, as this is the coefficient which determines the
magnitudes of the angular correlations.  Thus, the improvement
from 83\% left-handed helicity to 98\% spin up
in the antiproton basis translates into a factor of 1.45 increase
in the size of the correlations.
We plot the differential distributions
in top quark $p_T$ for the helicity and antiproton
bases as well as the total in Fig.~\ref{WStarPlot}.

%%%%%%%%%%%%%%%%%%%%%%%%%%%%%%%%%%%%%%%%%%%%%%%%%%%%%%%%%%%%%%%%
%%
%%      Wg FUSION PRODUCTION MECHANISM
%%
%%%%%%%%%%%%%%%%%%%%%%%%%%%%%%%%%%%%%%%%%%%%%%%%%%%%%%%%%%%%%%%%

\section{$W$-Gluon Fusion} \label{WgFusionSection}

The dominant production mechanism for single 
175 GeV top quarks at the Tevatron is the so-called
$Wg$ fusion process.  We consider the 
processes~(\ref{WgFusionProcess})
and hence the gauge-invariant
set of diagrams shown in Fig.~\ref{WgFusionDiagram}.
Once again we use the symbol for each particle to represent
its momentum.  For convenience, we employ the 
explicitly crossing-symmetric form in which all momenta
are outgoing.
Because $Wg$ fusion is a
$2\rightarrow 3$ process, the polarized production 
matrix elements squared 
for an arbitrary spin axis
are too complicated to reproduce here\cite{CarlsonThesis}.  
However, the sum over
all spins and colors 
may be simply written as\footnote{When the initial state partons
are chosen such that the $W$ momentum is timelike, one should add
the standard width term to the $W$ propagator.}
\beq
\vert\amp(0\rightarrow \bar{u}dgt\bar{b})\vert^2 =
{
{ g_W^4g_s^2\CKM N_c(N_c^2-1)}
\over
{ (2u\cdot d - m_W^2)^2 }%+ (m_W \Gamma_W)^2 }
}
\thinspace
\Bigl\vert{\cal Z}(u,d;t,b) + {\cal Z}(d,u;b,t) \Bigr\vert,
\eeq
where
\beq
{\cal Z}(u,d;t,b) = 
(2 d\cdot t) 
\Biggl\{
{ {t\cdot u} \over {t\cdot g} } 
- { { u\cdot(b{+}g) } \over {  b\cdot g} }
\biggl[
1 - { { b^2 } \over {  b\cdot g } } 
+ { {  t\cdot b} \over { t\cdot g} } 
\biggr]\Biggr\}.
\eeq
We present the relative contributions to the cross
section from each of the partonic initial states for this process
in Table~\ref{WgFusionPartons}.
%\footnote{We do not include the
%process $u\bar{d} \rightarrow t\bar{b}g$, which is infrared
%divergent.  In the soft and/or
%collinear gluon regime, it is a correction to the $W^{*}$
%production mechanism discussed in Sec.~\ref{WStarSection}.
%In the hard gluon regime, it may be viewed as a background
%to the $Wg$ fusion process, which may be reduced by exploiting
%its very different kinematics.
%Imposing a minimum gluon $p_T$ of 20 GeV,
%we find a total cross section of approximately 50 fb in this mode.}
As expected from the observation
that the proton contains two $u$ quarks while the antiproton
contains only one $\bar{d}$ quark, the $ug$ initial state
gives the largest contribution (74\%) to the total.
In terms of the helicity basis (defined in the zero momentum
frame of the incoming partons), we find that approximately
83\% of the top quarks have negative 
helicity (see Table~\ref{WgFusionPurities}), 
leaving significant room for improvement.

To illuminate our improved basis, we present the matrix
element squared for the production of spin down top quarks
in the basis where the spin axis is chosen to coincide with
the $d$ quark direction:
\beq
\vert\amp(0\rightarrow \bar{u}dgt_{\down}\bar{b})\vert^2 =
{
{ g_W^4 g_s^2 \CKM N_c(N_c^2-1)}
\over
{ (2u\cdot d - m_W^2)^2 }%+ (m_W \Gamma_W)^2 }
}
\thinspace
{
{ m_t^2 (g\cdot d)^2 }
\over
{ (t\cdot g)^2 }
}
\thinspace\Biggl\vert
{
{ u\cdot b }
\over
{ t\cdot d }
}\Biggr\vert.
\eeq
Besides being surprisingly simple, this result is significant
in that it comes exclusively from the lower diagram 
in Fig.~\ref{WgFusionDiagram};  hence, there are 
no inverse powers of $2b\cdot g$ from the $b$-quark propagator.
As is
well-known~\cite{WillenbrockDicus}, in the limit of vanishing
$b$-quark mass, the $Wg$ fusion process develops a collinear
singularity.  For the physical (non-zero) value of the $b$ mass, 
this is reflected in the tendency for the $b$ quark to be produced
at large pseudorapidity.  Thus, the majority of the total rate
comes from the regions of phase space where $2b\cdot g$ is small:  
hence the spin down component (no pole in $2b\cdot g$) is 
suppressed relative to the spin up component.   In fact, for
the $ug$ and $gu$ partonic initial states, we find that 
97\% of the tops are produced with spin up in this basis.

\begin{figure}[h]

\vspace*{16cm}
\includegraphics{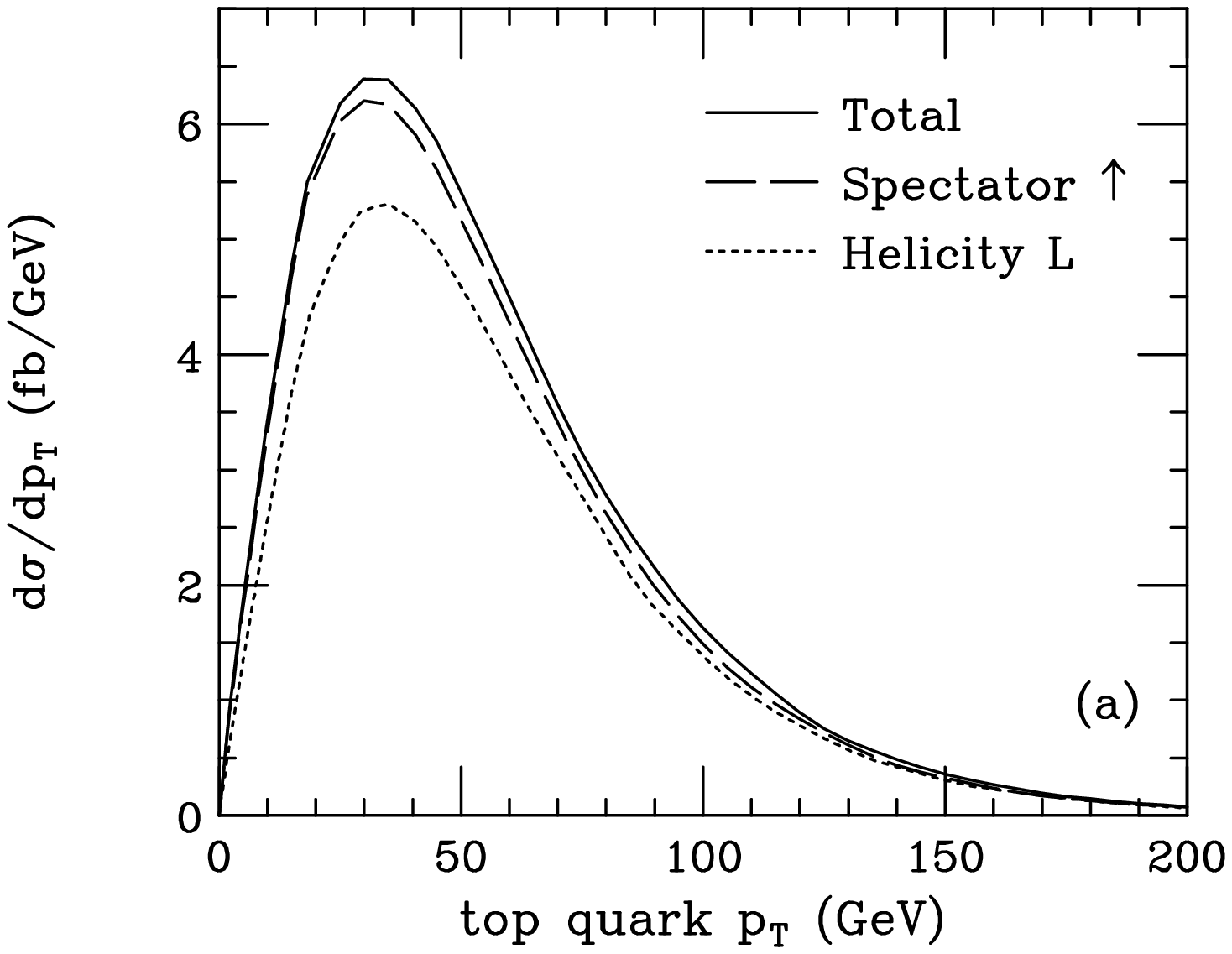}
\includegraphics{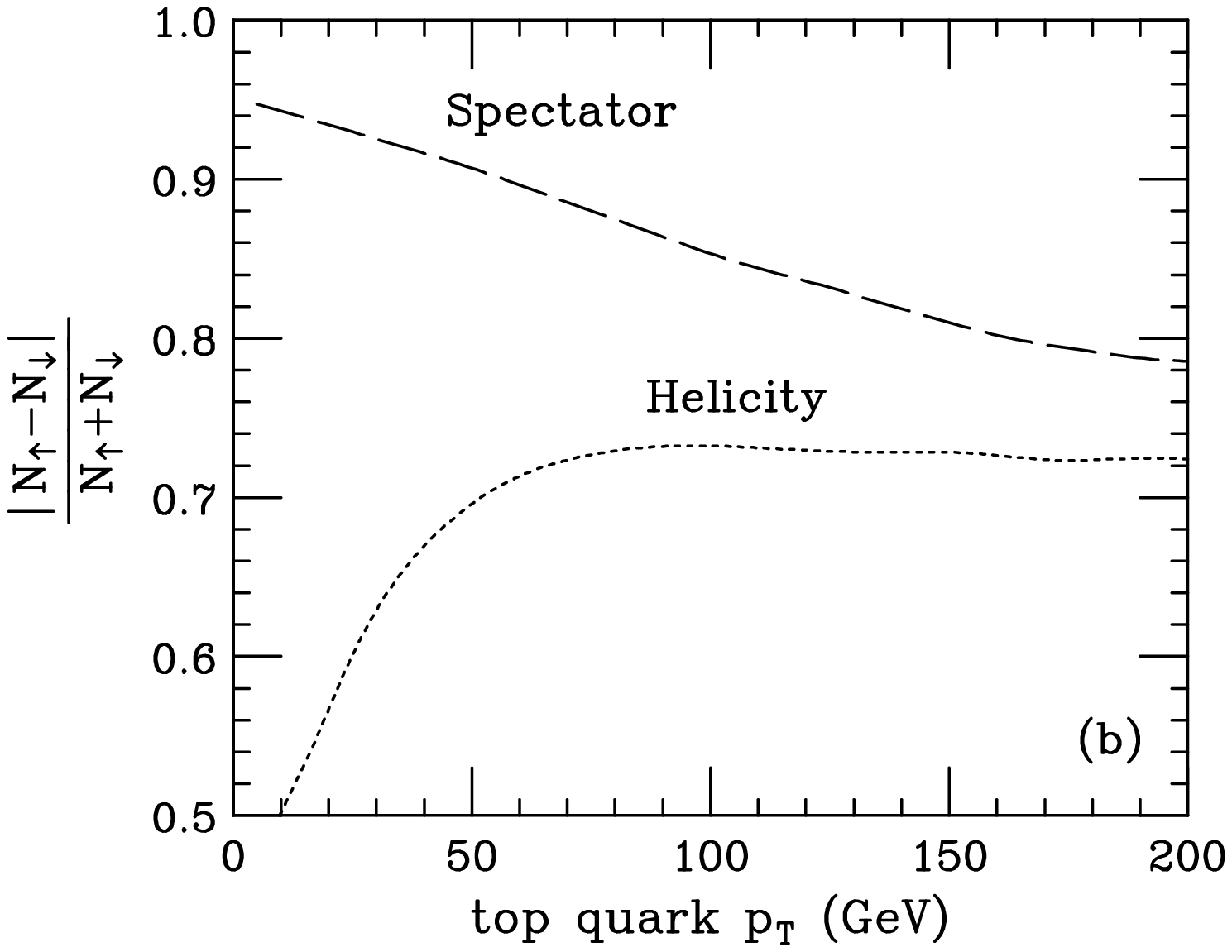}
\vspace{1.0cm}

%\caption[]{Figure for $Wg$ fusion mode.}
\caption[]{(a)  The differential cross sections
(total, spectator basis spin up, helicity basis left)
as a function
of the top quark transverse momentum for single top 
production via $Wg$ fusion at the Tevatron at 2.0 TeV.
(b)  The absolute value of the
spin asymmetry~(\ref{Asym}) plotted as a function of 
the top quark
transverse momentum for the helicity and spectator bases.
}
\label{WgFusionPlot}
\end{figure}

Since for the $ug$ and $gu$ initial states the $d$ quark
becomes the spectator jet, we define the ``spectator'' basis
by electing to use the direction of the spectator jet
(defined as the light jet appearing in the $\ell\nu b\bar{b}j$
final state) for the spin axis.  Although this picks the wrong
spin axis direction for the  
$g\bar{d}$ and $\bar{d}g$ initial states, it is correct the
majority of the time.  
We find that the  overall composition consists of 96\% spin up 
top quarks in this basis.
For comparison, we give the
results for the proton and antiproton bases 
in Table~\ref{WgFusionPurities}.  In terms of the spin
asymmetry defined in Eq.~(\ref{Asym}), we see that
the spectator basis represents a factor of 1.36 improvement
over the helicity basis.  The 
differential distributions
in top quark $p_T$ for the helicity and antiproton
bases as well as the total appear in Fig.~\ref{WgFusionPlot}.

%%%%%%%%%%%%%%%%%%%%%%%%%%%%%%%%%%%%%%%%%%%%%%%%%%%%%%%%%%%%%%%%
%%
%%      CONCLUSIONS
%%
%%%%%%%%%%%%%%%%%%%%%%%%%%%%%%%%%%%%%%%%%%%%%%%%%%%%%%%%%%%%%%%%

\section{Discussion and Conclusions} \label{CONCLUSIONS}

In this paper we have found that 
the direction of the $d$-type quark provides the most effective
spin axis for all single top production mechanisms.
However, experimentally we do not know with certainty which 
physical object comprises the $d$-type quark in a given event.
We have chosen the object which is most likely to be the $d$-type
quark.
In the case of the $W^{*}$ production
mechanism, this means the direction of the antiproton beam,
since it supplies the $\bar{d}$ quark 98\% of the time at the
Tevatron.  
Using the antiproton as our basis, we find that the top quark
is 98\% spin up.
As a result, the angular correlations
with this choice of spin axis are 45\% larger than 
those using the helicity basis.

For $Wg$ fusion, the situation is potentially more complicated.  
Nearly three-quarters of the cross section comes
from the situation where the proton donates a $u$ quark:  hence
the $d$ quark appears as the spectator jet in the final state.  
In double-tagged
events, identifying this jet is trivial;  in other cases, 
it may be necessary to assume that the jet with the largest
pseudorapidity is the spectator jet.
Although a full simulation 
is beyond the scope of this paper, it is clear 
that this identification can be achieved with a small
error rate because of the unique kinematics of this process.
Using the spectator jet as our basis, we find that
the top quark is 95\% spin up
and that the angular correlations are 36\% 
larger than the correlations using the helicity basis.

We have demonstrated that the helicity basis is {\it not} the
optimal basis for the discussion of angular correlations in
single top quark production at the Tevatron.  
Instead, we have shown that the direction of the $d$-type
quark provides a superior spin axis for {\it all}\ single
top production mechanisms.

%%%%%%%%%%%%%%%%%%%%%%%%%%%%%%%%%%%%%%%%%%%%%%%%%%%%%%%%%%%%%%%%
%%
%%      ACKNOWLEDGEMENTS
%%
%%%%%%%%%%%%%%%%%%%%%%%%%%%%%%%%%%%%%%%%%%%%%%%%%%%%%%%%%%%%%%%%

\acknowledgements

The Fermi National Accelerator 
Laboratory is operated by Universities Research Association,
Inc., under contract DE-AC02-76CHO3000 with the U.S. Department
of Energy. 
High energy physics research at the University of Michigan
is supported in part by the U.S. Department of Energy,
under contract DE-FG02-95ER40899.
GM would like to thank Tony Gherghetta for useful discussions
related to this work.
% The computer on which most of the monte carlos were generated 
% runs the Linux operating system. Thanks to Linus Torvalds 
% and his supporting cast of thousands for creating it.

%%%%%%%%%%%%%%%%%%%%%%%%%%%%%%%%%%%%%%%%%%%%%%%%%%%%%%%%%%%%%%%%
%%
%%      APPENDIX
%%
%%%%%%%%%%%%%%%%%%%%%%%%%%%%%%%%%%%%%%%%%%%%%%%%%%%%%%%%%%%%%%%%

\appendix
\section*{Angular Correlations in
$\noexpand\lowercase{u}\bar{\noexpand\lowercase{d}}\rightarrow
\noexpand\lowercase{t}\bar{\noexpand\lowercase{b}}\rightarrow
\bar\ell\nu\noexpand\lowercase{b}
\bar{\noexpand\lowercase{b}}$}
For the $W^*$ production process of 
single top quarks it is instructive 
to study the full matrix element including both 
the production and decay of the the top quarks
to see how the top spin correlation translates into
angular correlations in the events.
Consider the production of a top quark via
\beq
	u \bar{d} \rightarrow t \bar{b} \nonumber 
\eeq
and its subsequent semi-leptonic decay
\beq
	t  \rightarrow  \bar{\ell} \nu b. \nonumber
\eeq
Using the symbol of the particle to represent 
its momentum, the full matrix element squared for
this process including all correlations between
production and decay, summed over all colors
and spins, is given by
\beqa
\vert\amp(u \bar{d} \rightarrow t \bar{b} 
\rightarrow \bar{\ell} \nu b \bar{b})\vert^2
= && 2 N_c^2 g_W^8 \CKM
 (2u \cdot \bar{b}) \ts (2b \cdot \nu) 
  \ts\ts\Bigl\{2(t \cdot \bar{d})(t \cdot \bar{\ell}) 
  - t^2 \ts(\bar{d} \cdot \bar{\ell})\Bigr\}
\nonumber \\
 && \enspace\times\ts
 [(2u\cdot d-m_W^2)^2 + (m_W\Gamma_W)^2]^{-1}
  \ts\ts[(t^2-m_t^2)^2 + (m_t\Gamma_t)^2]^{-1}
\nonumber \\
 &&  \enspace\times\ts
  [(2\bar{\ell} \cdot \nu -m_W^2)^2 + (m_W\Gamma_W)^2]^{-1}
\label{WStarAll}
\eeqa
If we use the narrow width approximation for the top quark, then the
quantity in the curly brackets in Eq.~(\ref{WStarAll})
evaluated in the top quark rest frame is equal to 
\beq
m_t^2 \ts E_{\bar{d}} \ts  E_{\bar{\ell}} 
\ts(1+\cos \theta_{\bar{d}\bar{\ell}})
\eeq
where $E_i$ is the energy of the {\it i}\ts th particle and 
$\theta_{\bar{d}\bar{\ell}}$ is the angle between the 
$\bar{d}$-quark and the 
charged lepton in this frame. 
The $(1+\cos \theta_{\bar{d}\bar{\ell}})$ is precisely the 
correlation expected if the
top quark spin is along the direction of the $\bar{d}$-quark
momentum in the top quark rest frame.
This is confirmation of Eqs.~(\ref{WStarGenUP}) 
and (\ref{WStarGenDN})  and discussion that follows
in Section~\ref{WStarSection}. 

%%%%%%%%%%%%%%%%%%%%%%%%%%%%%%%%%%%%%%%%%%%%%%%%%%%%%%%%%%%%%%%%
%%
%%      REFERENCES
%%
%%%%%%%%%%%%%%%%%%%%%%%%%%%%%%%%%%%%%%%%%%%%%%%%%%%%%%%%%%%%%%%%

%%%%%%%%%%%%%%%%%%%%%%%%%%%%%%%%%%%%%%%%%%%%%%%%%%%%%%%%%%%%%%%%
%%
%%      FIGURE CAPTIONS
%%
%%%%%%%%%%%%%%%%%%%%%%%%%%%%%%%%%%%%%%%%%%%%%%%%%%%%%%%%%%%%%%%%
\vspace*{1cm}

%%%%%%%%%%%%%%%%%%%%%%%%%%%%%%%%%%%%%%%%%%%%%%%%%%%%%%%%%%%%%%%%
%%
%%      TABLES
%%
%%%%%%%%%%%%%%%%%%%%%%%%%%%%%%%%%%%%%%%%%%%%%%%%%%%%%%%%%%%%%%%%

\begin{table}
\caption{Fractional cross sections for single top production in the
$W^{*}$ channel at the Tevatron at 2.0 TeV, decomposed according
to the parton content of the initial state.  
We use the MRS(R1) structure functions~\protect\cite{structfun} 
evaluated at the scale $Q^2=m_W^2$.
We obtain a total cross section of approximately 0.33~pb.
%330~fb.
\label{WStarPartons}}
\begin{tabular}{ccccrcc}
&& $p$ & $\bar{p}$ & fraction && \\
\hline
&& $u$ & $\bar{d}$ & 98\%\enspace && \\
&& $\bar{d}$ & $u$ &  2\%\enspace &&
\end{tabular}
\end{table}

\begin{table}
\caption{Dominant spin fractions and asymmetries for the 
various bases studied for single top production in the  
$W^{*}$ channel at the Tevatron at 2.0 TeV.
\label{WStarPurities}}
\begin{tabular}{cccddcc}
&& basis & spin content & 
$\displaystyle{{{N_{\up}-N_{\down}}
\over{N_{\up}+N_{\down}}}}$ && \\[0.10in]
\hline
&& helicity   & 83\% $\downarrow$(L)             & $-$0.66 && \\
&& proton     & 83\% $\downarrow$\phantom{(L)}   & $-$0.67 && \\
&& antiproton & 98\% $\uparrow$\phantom{(L)}     &    0.96 && 
\end{tabular}
\end{table}

\begin{table}
\caption{Fractional cross sections for single top production in
the $Wg$ fusion channel at the Tevatron at 2.0 TeV, decomposed
according to the parton content of the initial state.
We use the MRS(R1) structure functions~\protect\cite{structfun}
evaluated at the scale $Q^2=m_W^2$. % and fixed $\alpha_s=0.1$.
We obtain a total cross section of approximately 0.47~pb.
%469~fb.
\label{WgFusionPartons}}
\begin{tabular}{ccccrcc}
&& $p$       & $\bar{p}$ &  fraction && \\
\hline
&& $u$       & $g$       & 74\%\enspace && \\
&& $g$       & $\bar{d}$ & 20\%\enspace && \\
&& $g$       & $u$       &  3\%\enspace && \\
&& $\bar{d}$ & $g$       &  3\%\enspace &&
\end{tabular}
\end{table}

\begin{table}
\caption{Dominant spin fractions and asymmetries for the 
various bases studied for single top production in the  
$Wg$ fusion channel at the Tevatron at 2.0 TeV.
\label{WgFusionPurities}}
\begin{tabular}{ccccdcc}
&& basis & spin content & 
$\displaystyle{{{N_{\up}-N_{\down}}
\over{N_{\up}+N_{\down}}}}$ && \\[0.10in]
\hline
&& helicity   & 83\% $\downarrow$(L)             & $-$0.67 && \\
&& proton     & 68\% $\uparrow$\phantom{(L)}     &    0.37 && \\
&& antiproton & 54\% $\downarrow$\phantom{(L)}   & $-$0.07 && \\
&& spectator  & 96\% $\uparrow$\phantom{(L)}     &    0.91 && 
\end{tabular}
\end{table}

\end{document}